\documentclass[aps,prd,nofootinbib,10pt,tightenlines,twocolumn,noti,floatfix,superscriptaddress,showkeys]{revtex4-2}

\usepackage[utf8]{inputenc}          
\usepackage{graphicx}         
\usepackage[usenames,dvipsnames]{xcolor}
\usepackage{array,dcolumn,longtable} 

\usepackage{comment}
\usepackage{amsmath,amssymb,amsfonts,slashed} 
\usepackage{mathtools}          
\usepackage[linktocpage,breaklinks]{hyperref}
\usepackage{mathrsfs}
\usepackage{txfonts}
\usepackage{bm}
\usepackage{stmaryrd}
\usepackage{tensor}
\usepackage[utf8]{inputenc}

\usepackage{epsfig}
\usepackage{epstopdf}

\usepackage{cleveref}

\definecolor{mred}{RGB}{127,0,25}
\definecolor{mdgr}{RGB}{51,51,51}
\definecolor{mag}{RGB}{211, 54, 130}
\definecolor{verm}{RGB}{164, 25, 0}

\hypersetup{colorlinks=true,
            citecolor=NavyBlue,
            linkcolor=NavyBlue,
            urlcolor=NavyBlue}
\usepackage{siunitx}                                  
\DeclareSIUnit{\fm}{\femto\metre}                     

\newcommand{\lie}{\mathscr{L}}

\newcount\hh
\newcount\mm
\mm=\time
\hh=\time
\divide\hh by 60
\divide\mm by 60
\multiply\mm by 60
\mm=-\mm
\advance\mm by \time
\def\hhmm{\number\hh:\ifnum\mm<10{}0\fi\number\mm}

\def\be{\begin{equation}}
\def\ee{\end{equation}}

\def\N{\mathcal{N}}

\begin{document}

\preprint{APS/123-QED}

\title{Local Hamiltonian dynamics from non-local action principles\\ and applications to binary systems in general relativity}

\author{Francisco M. Blanco}
\affiliation{Department of Physics, Cornell University, Ithaca, NY 14853, USA.}


\begin{abstract}

We consider a class of finite-dimensional dynamical systems whose equations of motion are derived from a non-local-in-time action principle. The action functional has a zeroth order piece derived from a local Hamiltonian and a perturbation in the form of a non-local functional of the trajectory on phase space. We prove that the dynamics of these systems admits a local Hamiltonian description to all order in the perturbation and we provide explicit formulae for the $\mathcal{N}^{\text{th}}$ order Hamiltonian and symplectic form in terms of the $(\mathcal{N}-1)^{\text{th}}$ order Hamiltonian flow. In the context of general relativity, these systems arise in the study of binary systems such as pairs of black holes or neutron stars in the small mass-ratio and post-Newtonian approximations. We provide applications of the formalism to binary systems in these regimes.

\end{abstract}

\maketitle

\vspace{0.1cm}

\section{Introduction}

In this paper we will study a class of finite dimensional dynamical systems with non-local-in-time interactions. Such systems can be described in terms of action functionals on paths in phase space, where the action contains multiple integrals with respect to time. Their equations of motion are integro-differential equations as opposed to the ordinary differential equations characteristic of Hamiltonian dynamical systems. A simple example of such an integro-differential equation is
\be
\ddot{x}(t)=f(x,t)+\int_{-\infty}^{\infty}K(t,t')x(t')dt'.
\ee
Here $f(x,t)$ is the local piece of the force and the integral is a non-local-in-time force that is a functional of the position $x(t')$. 

Non-local-in-time interactions generally arise when one "integrates out" some of the degrees of freedom of a system, giving rise to a non-local interaction between the remaining degrees of freedom. In the context of general relativity, in the study of binary systems such as gravitating pairs of black holes or neutron stars, non-local-in-time interactions appear in the small mass ratio and post-Newtonian approximations \cite{poissonwill,pound,4PNtailsDamour,blanchettails}. They are also useful for the description of cracks and other non-local deformations on materials \cite{SILLING2000175}. Non-local-in-time interactions are sometimes parametrized in terms of frequency dependent coefficients, such as the electric permittivity and susceptibility \cite{Jackson:1998nia}. They also appear in Fokker-Wheeler-Feynman electrodynamics \cite{wheelerfeynman}.

In the case of ordinary differential equations obtained from a Hamiltonian system, standard existence and uniqueness theorems \cite{Arnold} state that the space of solutions can be parametrized by initial data, i.e. points in phase space. When non-local-in-time interactions are included, however, it is not clear how to obtain a simple parametrization of the space of solutions \cite{llosa1994hamiltonian}. However, as is well known, when non-local-in-time interactions are treated perturbatively, the resulting dynamics can be cast as a local dynamical system, order by order. It is less well known, however, under what circumstances this local dynamical system admits a Hamiltonian description at each order. In this paper we derive the existence of such Hamiltonian description for a broad class of non-local-in-time action principles (Equation (\ref{eq:nonlocalaction}) below).

This paper is organized as follows: In section \ref{sec:nonlocalactions}, we review the dynamics obtained from non-local-in-time action principles and derive their equations of motion. We then treat the non-localities perturbatively to obtain local equations of motion order by order. In section \ref{sec:localham}, we prove that the local dynamics admits a local Hamiltonian description to any order in the perturbations. We provide explicit expressions for the Hamiltonian and symplectic form up to $\N^{\text{th}}$ order in terms of the $(\N-1)^{\text{th}}$ Hamiltonian flow. Sections \ref{sec:sf} and \ref{sec:pn} apply this result in the context of gravitational binary systems in general relativity. Uninterested readers can skip both sections altogether and focus on the rest of the paper. Section \ref{sec:pn} specializes to the dynamics of binary systems in the post-Newtonian approximation, where non-local effects start at fourth order \cite{4PNtailsDamour,blanchettails}. Section \ref{sec:sf} applies the results of this paper to extreme-mass-ratio inspirals, where the gravitational self interaction of a small object orbiting around a much larger one includes non-local effects due to the backscattering of gravitational waves \cite{poissonwill,pound}. In previous work \cite{Blanco:2022mgd,Blanco:2023jxf}, the conservative piece of the dynamics of a binary system in the small mass ratio regime was recast as a local Hamiltonian system to first order in the small mass-ratio. This paper generalizes the methods used in Refs. \cite{Blanco:2022mgd,Blanco:2023jxf} to a more general class of non-local systems and to arbitrary high orders in perturbation theory. It is our hope that they can be applied in other fields where these non-local-in-time interactions arise. 

This formalism is an extension of work done by Llosa and Vives in \cite{llosa1994hamiltonian}. The relation between their work and the results of this paper is discussed in appendix \ref{sec:appendix}.

\section{Dynamical systems described by non-local action principles}\label{sec:nonlocalactions}

We start this section by reviewing the description of phase space flows. Consider a phase space $\Gamma$ with coordinates
\be
Q^A=(q^\mu,p_\mu)
\ee
and a symplectic form $\Omega_0=\delta p_\mu \wedge \delta q^\mu$. We define a flow in phase space $X_s(Q):\mathbb{R}\times\Gamma\rightarrow \Gamma$ which takes any point $Q\in\Gamma$ into $X_s(Q)\in \Gamma$. The flow is required to be the identity map at $s=0$
\be
X_0(Q)=Q
\ee
and to satisfy the composition rule
\be\label{eq:propcomp}
X_s(X_{s'}(Q))=X_{s+s'}(Q)
\ee
for all $s,s'\in \mathbb{R}$. A flow $X_s(Q)$ on phase space will be determined by a vector field $\vec{V}=V^A \partial_A$ according to
\begin{equation}\label{eq:hamflow}
\frac{dX_s^A(Q)}{ds}=V^A[X_s(Q)].
\end{equation}
If we specialize equation (\ref{eq:hamflow}) to $s=0$ we get
\be\label{eq:hamflowzer0}
\left.\frac{d X^A_s(Q)}{ds}\right|_{s=0}=V^A(Q)
\ee
so the flow is determined by its derivative at $s=0$. Throughout this paper, we will parametrize and characterize flows by their derivatives (\ref{eq:hamflowzer0}) at $s=0$ with the understanding that the full flows are obtained by solving equation (\ref{eq:hamflow}).

We will consider dynamical systems described by non-local action functionals of paths $X_s$ of the form
\be\label{eq:action}
S[X]=\int p_\mu dq^\mu-\int H_0(X_s)ds+S_{nl}[X].
\ee
Here, $H_0 (Q)$ is a local Hamiltonian function on phase space and the non-local piece of the action is
\be\label{eq:nonlocalaction}
\begin{aligned}
S_{nl}[X]&=-\sum_{n=2}^N \frac{\epsilon_n}{n}\int ds_1 \dots ds_n \ \times \\
&\times \mathcal{G}_n(X_{s_1},\dots,X_{s_n};s_2-s_1,\dots,s_n-s_1),
\end{aligned}
\ee
where $\mathcal{G}_n$ is some n-point function $\mathcal{G}_n: \Gamma^{n}\times \mathbb{R}^{n-1}\rightarrow \mathbb{R}$. Here $\epsilon_n$ is a formal expansion parameter used to keep track of orders in the non-local action and $N$ is a finite but otherwise arbitrary positive integer. Note that because the n-point function $\mathcal{G}_n$ is integrated n times, the non-local action will automatically pick out its fully symmetric piece, so that without loss of generality we can assume that $\mathcal{G}_n$ satisfies
\be\label{eq:symprop}
\begin{aligned}
\mathcal{G}_n(X_{s_1},\dots,X_{s_n};\sigma_{12},\dots,&\sigma_{1n})=\\
&\mathcal{G}_n(X_{s_{p_1}},\dots,X_{s_{p_n}};\sigma_{p_1p_2},\dots,\sigma_{p_1p_n})
\end{aligned}
\ee
for all $(s_1,\dots,s_n)$. Here $\sigma_{ij}=s_j-s_i$ for short and $\{p_i\}$ is any permutation of the integers from 1 to $n$. We will also assume that the n-point functions satisfy asymptotic fall-off conditions given in detail in equation (\ref{eq:propGdecay}) below.

We will write the equations of motion in terms of a function $\Phi(Q,Q',[X])$ which is a local function of two points $Q$ and $Q'$ in phase space in its first two arguments and a functional of a trajectory $X_s$ which passes through $Q'$ at $s=0$ in its last argument. The definition of $\Phi$ is
\be\label{eq:phi}
\begin{aligned}
\Phi(Q,Q',[X])&=\sum_{n=2}^N\epsilon_n \int ds_2 \dots ds_n \ \times \\
& \times \mathcal{G}_n\big(Q,X_{s_2}(Q'),\dots,X_{s_n}(Q');s_2,\dots,s_n\big).
\end{aligned}
\ee
The equations of motion are obtained by varying the action functional (\ref{eq:action}) with respect to the trajectory $X$. The variation of the nth term in the non-local piece will give $n$ contributions with derivatives acting on each of the first $n$ arguments of $\mathcal{G}_n$. From property (\ref{eq:symprop}), it follows that all these contributions coincide, so we can add them up. The final result is a factor of $n$ times the derivative with respect to the first argument of $\Phi$. The resulting equations of motion are
\be\label{eq:hameqs}
\Omega^0_{AB}\left.\frac{d X^B_s}{ds}\right.=\left[\frac{\partial}{\partial Q^A}H_{0}(Q)+\frac{\partial}{\partial Q^A}\Phi(Q,Q',[X])\right]_{Q'=Q=X_s}.
\ee
Here the subscript $Q'=Q$ means that first two arguments of $\Phi$ are evaluated at coincidence after differentiating $\Phi$ with respect to its first entry. Subsequently, the whole right hand side is evaluated at $X_s$.

\subsection{Local dynamical systems obtained by treating non-localities perturbatively}\label{sec:orderred}

Equation (\ref{eq:hameqs}) is an integro-differential system of equations for the trajectories $X_s$ on phase space, as opposed to a differential system of equations that depend locally on a point $Q$, as is the case for Hamilton's equations derived from action principles without non-localities. Because of this property, solutions will generally not be parametrized by initial data $Q$. In fact, the space of initial data required to determine solutions of integro-differential systems of equations can be, in general, infinite dimensional and require derivatives of $x^\mu$ and $p_\mu$ with respect to time of all orders \cite{llosa1994hamiltonian}.

However, if we take the non-local contribution to the action to be small we can treat the problem perturbatively. We define a sequence of phase space flows $\bar{X}^{(\N)}_s(Q)$ by induction as follows. The zeroth order flow $\bar{X}^{(0)}_s(Q)$ is generated by the Hamiltonian $H_0$, with all non-local terms in equation (\ref{eq:hameqs}) dropped. Then, we can evaluate the functional dependence of $\Phi$ in equation (\ref{eq:hameqs}) on the zeroth order flow and define the first order flow $\bar{X}^{(1)}_s(Q)$ by
\be
\Omega^0_{AB}\left.\frac{d \bar{X}_s^{(1)B}(Q)}{ds}\right|_{s=0}=\frac{\partial}{\partial Q^A}H_{0}+\left[\frac{\partial}{\partial Q^A}\Phi(Q,Q',[\bar{X}^{(0)}])\right]_{Q'=Q}.
\ee
This process can be repeated to any desired order to define the $\N^{\text{th}}$ order flow in terms of the $(\N-1)^{\text{th}}$ flow as
\be\label{eq:hameqsnth}
\Omega^0_{AB}\left.\frac{d\bar{X}^{(\N)B}_s(Q)}{ds}\right|_{s=0}=\frac{\partial}{\partial Q^A}H_{0}+\left[\frac{\partial}{\partial Q^A}\Phi(Q,Q',[\bar{X}^{(\N-1)}])\right]_{Q'=Q}.
\ee
Equation (\ref{eq:hameqsnth}) is a set of ordinary differential equation which determines the $\N^{\text{th}}$ flow, once the $(\N-1)^{\text{th}}$ flow is specified\footnote{Equations of motion like these can be derived from a pseudo-Hamiltonian function, first defined in reference \cite{Blanco:2022mgd}. We detail the relation of this paper to pseudo-Hamiltonians in appendix \ref{sec:appendix2}, although we will not use that formalism here.}. Hence all the flows are determined by induction\footnote{As  is well known, perturbative expansions of this form can break down after long timescales when there are dissipative effects present. Here, we are concerned only with conservative dynamics and so we can neglect this issue.}.

The flow determined by equation (\ref{eq:hameqsnth}) agrees with the exact flow determined by equation (\ref{eq:hameqs}) up to corrections of order $O(\epsilon_1^{q_1}\times\epsilon_2^{q_2}\times\dots \times\epsilon_N^{q_N})$ with $\sum_{n=1}^N q_i=\N+1$. For simplicity, when we expand the Hamiltonian and symplectic form explicitly bellow, we will introduce a formal expansion parameter $\epsilon$ such that 
\be\label{eq:epsilon}
O(\epsilon^{\N+1})\equiv O(\epsilon_1^{q_1}\times\epsilon_2^{q_2}\times\dots \times\epsilon_N^{q_N}) \ ,\ \ \sum_{n=1}^N q_i=\N+1. 
\ee
We will also assume that the sequence of flows $\bar{X}^{(\N)}_s(Q)$ are such that the n-point functions $\mathcal{G}_n$ introduced in the non-local action principle in equation (\ref{eq:nonlocalaction}) satisfy the following property: For any $j\in[1,n]$ and with all $s_k$ with $k\neq j$ fixed, the limit when $s_j\rightarrow \pm\infty$ of the n-point function $\mathcal{G}_n$ evaluated on the flow $\bar{X}^{(\N)}_s$ is zero
\be\label{eq:propGdecay}
\begin{aligned}
\lim_{s_j\rightarrow \pm\infty}\mathcal{G}_n&\big(\bar{X}^{(\N)}_{s_1},\dots,\bar{X}^{(\N)}_{s_j},\dots,\bar{X}^{(\N)}_{s_n};\\
&s_2-s_1,\dots,s_j-s_1,\dots,s_n-s_1\big)=0.
\end{aligned}
\ee

\section{Local Hamiltonian description}\label{sec:localham}

In this section, we will obtain a local Hamiltonian description for the $\N^{\text{th}}$ order flow, in term of the known $(\N-1)^{\text{th}}$ order flow. We  define the Hamiltonian and symplectic form in this subsection and derive their equivalence to the system (\ref{eq:hameqsnth}) in the next subsection. 

Given a phase space flow $X_s(Q)$ and a point $Q$ in $\Gamma$, we define a function
\be
\begin{aligned}\label{eq:defpsi0}
\Psi(Q,[X])&=\frac{1}{2}\sum_{n=2}^N\epsilon_n\int ds_1 \dots ds_n \  \chi(s_1,\dots,s_n) \times \\
& \times \frac{\partial}{\partial s_1} \mathcal{G}(X_{s_1}(Q)\dots,X_{s_n}(Q);s_2-s_1,\dots,s_n-s_1)
\end{aligned}
\ee
where
\be
\chi(s_1,\dots,s_n)=\frac{\text{sgn}(s_1)-\text{sgn}(s_2)-\dots-\text{sgn}(s_n)}{2}.
\ee
Here the partial derivative $\partial/\partial s_1$ indicates that the derivative acts only on the explicit dependence of the n-point function in its last $n-1$ arguments and not on the implicit dependence that arises through $X_{s_1}$.

We now define the local Hamiltonian function in terms of the $(\N-1)^{\text{th}}$ flow as
\begin{equation}\label{eq:Hdef}
    H^{(\N)}(Q)=H_0(Q)+\Phi^{(\N)}(Q)+\Psi^{(\N)}(Q),
\end{equation}
where
\begin{subequations}\label{eq:defpsi}
\begin{align}
    \Phi^{(\N)}(Q)&=\Phi(Q,Q,[\bar{X}^{(\N-1)}]),\\
    \Psi^{(\N)}(Q)&=\Psi(Q,[\bar{X}^{(\N-1)}]).
\end{align}
\end{subequations}
We also define a new function of $n$ points on phase space as
\be
\begin{aligned}\label{eq:Kdef}
K_n^{(\N)}(Q_1,&\dots,Q_n)=\epsilon_n\int ds_1 \dots ds_n\ \chi(s_1,\dots,s_n)\ \times \\
&\times \mathcal{G}_n\big(\bar{X}^{(\N-1)}_{s_1}(Q_1),\dots,\bar{X}^{(\N-1)}_{s_n}(Q_n);s_2-s_1,\dots,s_n-s_1\big).
\end{aligned}
\ee
Note that the subscript $n$ labels the number of arguments in the n-point function $\mathcal{G}_n$ while the superscript $(\N)$ denotes an object constructed from the $(\N-1)^{\text{th}}$ order flow and contains terms of order $O(\epsilon^\N)$ and lower.
Using definition (\ref{eq:Kdef}) we define the local symplectic form
\begin{subequations}\label{eq:symdef}
\begin{align}
\Omega^{(\N)}&=\Omega_0+\Delta\Omega^{(\N)},\\ 
\Omega_0&=\delta p_\mu \wedge \delta q^\mu,\\
\Delta \Omega^{(\N)}_{AB}(Q)&=\left[ \sum_{n=2}^N \sum_{m=2}^n\frac{\partial^2}{\partial Q_1^{[A}\partial Q_m^{B]}} K_n^{(\N)}(Q_1,\dots,Q_n)\right]_{\{Q_j\}=Q}\label{eq:symdef3}
\end{align}
\end{subequations}
where $\{Q_j\}=Q$ means that we evaluate at coincidence $Q_1=Q_2=\dots=Q_n=Q$. Here, brackets denote antisymmetrization $\Omega_{[AB]}=\frac{1}{2}\big(\Omega_{AB}-\Omega_{BA}\big)$. 

Both $H^{(\N)}$ and $\Omega^{(\N)}$ can be expanded perturbatively using the formal expansion parameter (\ref{eq:epsilon}) as
\begin{subequations}
    \begin{align}
        H^{(\N)}&=H_0+\sum_{r=1}^{\N}\epsilon^r H^{[r]}\\
        \Omega^{(\N)}&=\Omega_0+\sum_{r=1}^\N \epsilon^r \Delta \Omega^{[r]}
    \end{align}
\end{subequations}
where a superscript $[r]$ indicates a term that is exclusively $O(\epsilon^r)$, as opposed to a superscript $(\N)$ which indicates a term that contains contributions of order $O(\epsilon^\N)$ and lower. 

\subsection{Derivation of Hamilton formulation}\label{sec:localderiv}

In this subsection we will prove that the Hamiltonian function (\ref{eq:Hdef}) equipped with the symplectic form (\ref{eq:symdef}) reproduces the perturbative local dynamical system (\ref{eq:hameqsnth}) up to corrections of order $O(\epsilon^{\N+1})$.

The Hamiltonian function (\ref{eq:Hdef}) equipped with the symplectic form (\ref{eq:symdef}) determines the flow
\be
\begin{aligned}\label{eq:hameqs1}
\Big[\Omega^0_{AB}+\Delta \Omega^{(\N)}_{AB}\Big]\left.\frac{d\bar{X}^{(\N)}_s}{ds}\right|_{s=0}=\frac{\partial}{\partial Q^A}\Big[H_0+ \Phi^{(\N)}+ \Psi^{(\N)}\Big].
\end{aligned}
\ee
First, note that since we want the equations of motion to be accurate up to corrections of order $O(\epsilon^{\N+1})$, we can drop higher order corrections in the second term in the left side of equation (\ref{eq:hameqs1})
\be\label{eq:symcontrred}
\Delta \Omega^{(\N)}_{AB}\left.\frac{d\bar{X}^{(\N)}_s}{ds}\right|_{s=0}=\Delta \Omega^{(\N)}_{AB}\left.\frac{d\bar{X}^{(\N-1)}_s}{ds}\right|_{s=0}+O(\epsilon^{N+1})
\ee
where we replaced $\bar{X}^{(\N)}$ with $\bar{X}^{(\N-1)}$ since $\Delta \Omega^{(\N)}$ is $O(\epsilon)$. We will calculate the first term in the right hand side of equation (\ref{eq:symcontrred}) in a series of steps. First, the contraction $\Delta \Omega^{(\N)}_{AB}\left.d \bar{X}^{(\N-1)}_s/ds\right|_{s=0}$ will have two pieces coming from the antisymetrization of the indices $AB$ in equation (\ref{eq:symdef3}). The first one is  
\be\label{eq:step0}
\left.\frac{d\bar{X}^{(\N-1)B}_s}{ds}\right|_{s=0}\times \left[\frac{\partial^2}{\partial Q_1^{A}\partial Q^{B}_m}K_n^{(\N)}(Q_1,\dots,Q_n)\right]_{\{Q_j\}=Q}.
\ee
The derivative of $\bar{X}^{(\N-1)B}_s$ is evaluated at $Q$ but we are allowed to move it inside the brackets and evaluate it at $Q_m$, since the bracket is evaluated at coincidence $\{Q_j\}=Q$. Using property (\ref{eq:propcomp}), the contraction $\left.d\bar{X}^B_s/ds\right|_{s=0}\ \partial/\partial Q^{B}$ acting on any function $f(\bar{X}_s(Q),s)$ will create a total derivative $d/ds$, minus a correction $\partial/\partial s$ due to the explicit time dependence of $f$ \begin{equation}\label{eq:idcontr}
    \begin{aligned}
\frac{d \bar{X}^A}{ds} \frac{\partial}{\partial Q^A}f\big(\bar{X}_s(Q),s\big)&=\left.\frac{d}{d\Delta s}\right|_{\Delta s=0}f\big(\bar{X}_{s+\Delta s}(Q),s\big)\\
&=\left[\frac{d}{ds}-\frac{\partial}{\partial s}\right] f\big(\bar{X}_{s}(Q),s\big).
\end{aligned}
\end{equation}
Using the identity (\ref{eq:idcontr}) in equation (\ref{eq:step0}) we get
\be\label{eq:totalandpartialder}
\begin{aligned}
&\Big\{\frac{\partial}{\partial Q_1^A}\int ds_1 \dots ds_n\ \chi(s_1,\dots,s_n)\big[\frac{d}{ds_m}-\frac{\partial}{\partial s_m}\big]\times \\
&\times  \mathcal{G}_n\big(\bar{X}^{(\N-1)}_{s_1}(Q_1),\dots,\bar{X}^{(\N-1)}_{s_n}(Q_n);s_2-s_1,\dots,s_n-s_1\big) \Big\}_{\{Q_j\}=Q}.
\end{aligned}
\ee
We integrate by parts the total derivative $d/ds_m$, use property (\ref{eq:propGdecay}) to throw away boundary terms, relabel $s_m\leftrightarrow s_1$ and use the properties (\ref{eq:symprop}) and 
\be\label{eq:propchi}
\frac{d}{ds_j}\chi(s_1,\dots,s_n)=\begin{cases}&+\delta(s_1) \ \ \text{for }j=1\\
&-\delta(s_j)\ \ \text{for }j=2,\dots,n\end{cases}
\ee
to get
\be\label{eq:step1}
\begin{aligned}
\Big[\frac{\partial}{\partial Q_m^A}&\int ds_2 \dots ds_n \times \\
&\times \mathcal{G}_n(Q_1,\bar{X}^{(\N-1)}_{s_2}(Q_2)\dots,\bar{X}^{(\N-1)}_{s_n}(Q_n);s_2,\dots,s_n)\Big]_{\{Q_j\}=Q}.
\end{aligned}
\ee
Note that once we sum over $n$ and $m$ this term will give the $n-1$ last derivatives of the n-point function $\mathcal{G}_n$ in equation (\ref{eq:phi}), which can be expressed as
\be
\left[\frac{\partial}{\partial Q^{A'}}\Phi(Q,Q',[\bar{X}^{(\N-1)}])\right]_{Q'=Q},
\ee
where the prime index in $Q^{A'}$ means that the derivative acts on $Q'$ but not $Q$.
Regarding the term proportional to the partial derivative $\partial/\partial s_m$in equation (\ref{eq:totalandpartialder}), note that we can pull the sum $\sum_{m=2}^n$ from equation (\ref{eq:symdef3}) inside $K$ to get
\be
\begin{aligned}\label{eq:step2}
&\Big[\frac{\partial}{\partial Q_1^A}\int ds_1 \dots ds_n\ \chi(s_1,\dots,s_n)\Big[-\sum_{m=2}^{n}\frac{\partial}{\partial s_m}\Big]\times \\
&\times  \mathcal{G}_n\big(\bar{X}^{(\N-1)}_{s_1}(Q_1),\dots,\bar{X}^{(\N-1)}_{s_n}(Q_n);s_2-s_1,\dots,s_n-s_1\big) \Big]_{\{Q_j\}=Q}\\
&=\Big[\frac{\partial}{\partial Q_1^A}\int ds_1 \dots ds_n\ \chi(s_1,\dots,s_n)\frac{\partial}{\partial s_1}\times\\
&\times  \mathcal{G}_n\big(\bar{X}^{(\N-1)}_{s_1}(Q_1),\dots,\bar{X}^{(\N-1)}_{s_n}(Q_n);s_2-s_1,\dots,s_n-s_1\big) \Big]_{\{Q_j\}=Q}
\end{aligned}
\ee
where we used a chain rule to replace the derivatives respect to all the $s_m$ with a derivative respect to $s_1$.

Now, we move on to the other piece of the contraction $\Delta \Omega^{(\N)}_{AB}\left.d\bar{X}^{(\N-1)}_s/ds\right|_{s=0}$ coming from the antisymmetrization of indices $AB$ in equation (\ref{eq:symdef3})
\be
-\left.\frac{d\bar{X}_s^{(\N-1)B}}{ds}\right|_{s=0}\times \left[\frac{\partial^2}{\partial Q_m^{A}\partial Q^{B}_1}K_n^{(\N)}(Q_1,\dots,Q_n)\right]_{\{Q_j\}=Q}.
\ee
The contraction once again will give a total derivative $d/ds_1$ minus a correction $\partial/\partial s_1$ due to the explicit time dependence on $K$. Integrating by parts the total derivative recovers equation (\ref{eq:step1}). The term proportional to the partial derivative is
\be
\begin{aligned}\label{eq:step4}
&\Big[\frac{\partial}{\partial Q_m^A}\int ds_1 \dots ds_n\ \chi(s_1,\dots,s_n)\frac{\partial}{\partial s_1}\times \\
&\times  \mathcal{G}_n\big(\bar{X}^{(\N-1)}_{s_1}(Q_1),\dots,\bar{X}^{(\N-1)}_{s_n}(Q_n);s_2-s_1,\dots,s_n-s_1\big) \Big]_{Q_n=\dots=Q}.
\end{aligned}
\ee
Now we apply the sum over $m$ to this last term and add it to the term in equation (\ref{eq:step2}) to create a derivative $\partial_A$ acting on every argument of $K$. Putting equations (\ref{eq:step1}), (\ref{eq:step2}) and (\ref{eq:step4}) together we get that
\be\label{eq:step5}
\begin{aligned}
\Delta \Omega^{(\N)}_{AB}\left.\frac{d\bar{X}^{(\N)}_s}{ds}\right|_{s=0}&=\left[\frac{\partial}{\partial Q^{A'}} \Phi(Q,Q',[\bar{X}^{(\N-1)}])\right]_{Q'=Q}\\
&+\partial_A \Psi^{(\N)}(Q)+O(\epsilon^{\N+1}).
\end{aligned}
\ee
Plugging this into equation (\ref{eq:hameqs1}) we see that the first term on the right hand side of equation (\ref{eq:step5}) cancels all the extra derivatives respect to the last (n-1) arguments of the n-point function in $\partial_A \Phi$ in equation (\ref{eq:hameqs1}). The last term in the right hand side of equation (\ref{eq:step5}) cancels the term $\partial_A \Psi^{(\N)}$ in equation (\ref{eq:hameqs1}) and we recover equation (\ref{eq:hameqsnth}) up to corrections of order $O\big(\epsilon^{\N+1})$
as desired. 

\section{Alternative formulation of local Hamiltonian system}\label{sec:diffeo}

In this section we prove that, up to any order in $\epsilon_1$, $\epsilon_2$, $\dots,$ $\epsilon_n$, there exists a diffeomorphism in phase space that puts the symplectic form (\ref{eq:symdef}) in canonical form. We then apply this result up to second order and give explicit expressions for the diffeomorphism and the resulting Hamiltonian. We use arrows $\vec{V}=V^A \partial_A$ for vectors and tildes $\tilde{\omega}=\omega_A dQ^A$ for 1-forms. Indices will be raised and  lowered by contraction with the first index on the zeroth order symplectic form $\Omega_{AB}$. 

We consider a one-parameter family of diffeomorphisms $\varphi(\epsilon):\Gamma\rightarrow \Gamma$ that transform the $\N^{\text{th}}$ order Hamiltonian system $(\Omega_0+\Delta\Omega^{(\N)},H^{(\N)})$ to an equivalent Hamiltonian system 
\be\label{eq:transformedhamsys}
(\varphi_* \Omega_0 +\varphi_*\Delta\Omega^{(\N)},\varphi_*H^{(\N)})
\ee
where $\varphi_*$ is the pullback\footnote{As is well known, these transformations can be seen from a passive or an active viewpoint. The passive viewpoint considers the transformation to be a coordinate transformation, keeping all fields fixed. The active viewpoint considers the transformation as a field redefinition instead, with all coordinates unchanged. Both viewpoints are equivalent, but in this paper we adopt the active viewpoint for clarity.} defined by the diffeomorphism $\varphi(\epsilon)$. 

We now specialize the diffeomorphism $\varphi$ to make the new Hamiltonian system take the form $(\Omega_0,\varphi_* H^{(\N)})$, i.e. to make the transformed symplectic form coincide with the original zeroth order symplectic form 
\be\label{eq:symtocan}
\varphi_*\Omega^{(\N)}=\Omega_0+O(\epsilon^{\N+1}).
\ee
First, note that we can express the perturbation (\ref{eq:symdef}) to the symplectic form $\Delta \Omega^{(\N)}$ as an exact form. We define the 1-form
\be\label{eq:defxi}
\xi_A^{(\N)}=-\frac{1}{2}\sum_{n=2}^N \Big[\frac{\partial}{\partial Q_1^A}K^{(\N)}_n(Q_1,\dots,Q_n)\Big]_{\{Q_j\}=Q}
\ee
such that the perturbation to the symplectic form is
\be\label{eq:symisexact}
\begin{aligned}
\Delta \Omega^{(\N)}_{AB}(Q)&=\big(d\tilde{\xi}^{(\N)}\big)_{AB}\\
&= \partial_A \xi^{(\N)}_B-\partial_B \xi^{(\N)}_A.
\end{aligned}
\ee
The 1-form $\tilde{\xi}^{(\N)}$ is sourced by the $(\N-1)^{\text{th}}$ order flow and is accurate up to corrections of order $O(\epsilon^{\N+1})$. 
Plugging equation (\ref{eq:symisexact}) into the symplectic form (\ref{eq:symdef}) we get 
\be\label{eq:sym1form}
\Omega^{(\N)}=\Omega_0+ d\tilde{\xi}^{(\N)} +O(\epsilon^{\N+1}).
\ee 
Now, consider a one-parameter family of diffemorphisms $\varphi(\epsilon):\Gamma\rightarrow \Gamma$. We parametrize this diffeomorphism up to order $\N$ by $\N$ vector fields $\vec{\zeta}_i$ with $i=1,\dots,\N$ as
\be\label{eq:diffeo}
\varphi(\epsilon)=\mathcal{D}_{\vec{\zeta}_\N}(\epsilon^\N)\circ \mathcal{D}_{\vec{\zeta}_{\N-1}}(\epsilon^{\N-1})\circ\dots\circ \mathcal{D}_{\vec{\zeta}_1}(\epsilon)\left[1+O(\epsilon^{\N+1})\right],
\ee
where the mapping $\mathcal{D}_{\vec{\zeta}}(\epsilon)$ moves any point $\epsilon$ units along the vector field $\vec{\zeta}$. The pullback $\varphi_*$ can be expressed in terms of Lie derivatives as
\be\label{eq:pertpullback}
\begin{aligned}
\varphi_*&=1+\epsilon \lie_{\vec{\zeta}_1}+\frac{\epsilon^2}{2}\lie_{\vec{\zeta}_1}\lie_{\vec{\zeta}_1}+\frac{\epsilon^3}{6}\lie_{\vec{\zeta}_1}\lie_{\vec{\zeta}_1}\lie_{\vec{\zeta}_1}\\
&+\epsilon^2 \lie_{\vec{\zeta}_2}+\epsilon^3\lie_{\vec{\zeta}_1}\lie_{\vec{\zeta}_2}\\
&+\epsilon^3\lie_{\vec{\zeta}_3}+\dots
\end{aligned}
\ee
We want this diffeomorphism to make the symplectic form coincide with $\Omega_0$ up to $\N^{\text{th}}$ order, as in equation (\ref{eq:symtocan}). Combining equations (\ref{eq:symtocan}) and (\ref{eq:pertpullback}) and inverting the pullback perturbatively, we can invert equation (\ref{eq:symtocan}) to get
\be\label{eq:syminvert}
\begin{aligned}
\Omega^{(\N)}&=\Omega_0-\epsilon \mathscr{L}_{\vec{\zeta}_1}\Omega_0+\frac{\epsilon^2}{2}\lie_{\vec{\zeta}_1}\lie_{\vec{\zeta}_1}\Omega_0-\frac{\epsilon^3}{6}\lie_{\vec{\zeta}_1}\lie_{\vec{\zeta}_1}\lie_{\vec{\zeta}_1}\Omega_0\\
&-\epsilon^3  \mathscr{L}_{\vec{\zeta}_2}\Omega_0+\epsilon^3 \lie_{\vec{\zeta}_2}\lie_{\vec{\zeta}_1}\Omega_0 -\epsilon^3 \lie_{\vec{\zeta}_3}\Omega_0+\dots
\end{aligned}
\ee
We expand the 1-form (\ref{eq:defxi}) in powers of the formal parameter $\epsilon$ defined in equation (\ref{eq:epsilon})
\be\label{eq:1formorders}
\tilde{\xi}^{(\N)}=\sum_{r=1}^{\N}\epsilon ^r \tilde{\xi}^{[r]}
\ee
where $\tilde{\xi}^{[r]}$ is the piece of $\tilde{\xi}^{(\N)}$ of order $O(\epsilon^{r})$ and can be obtained by expanding the flow $\bar{X}^{(\N-1)}_s(Q)$ in the definition (\ref{eq:Kdef}) and plugging the expansion back into equation (\ref{eq:defxi}).
We plug the expansion (\ref{eq:1formorders}) into (\ref{eq:sym1form}) and then into equation (\ref{eq:syminvert}) and equate coefficients of powers of $\epsilon$ on both sides to obtain
\begin{subequations}\label{eq:allconds}
    \begin{align}    \label{eq:1stcond}
    \mathscr{L}_{\vec{\zeta}_1}\Omega_0&=-d\tilde{\xi}^{[1]},  \\    \label{eq:2ndcond}
        \mathscr{L}_{\vec{\zeta}_2}\Omega_0&=-d\tilde{\xi}^{[2]}+\frac{1}{2}\mathscr{L}_{\vec{\zeta}_1}\mathscr{L}_{\vec{\zeta}_1}\Omega_0,\\ \label{eq:3rdcond}
        \mathscr{L}_{\vec{\zeta}_3}\Omega_0&=-d\tilde{\xi}^{[3]}+\mathscr{L}_{\vec{\zeta}_2}\mathscr{L}_{\vec{\zeta}_1}\Omega_0-\frac{1}{6}\mathscr{L}_{\vec{\zeta}_1}\mathscr{L}_{\vec{\zeta}_1}\mathscr{L}_{\vec{\zeta}_1}\Omega_0,\\
        &\vdots \nonumber
    \end{align}
\end{subequations}
Using Cartan's magic formula and the fact that the symplectic form $\Omega_0$ is closed, we can prove that the Lie derivative of the zeroth order symplectic form $\Omega_0$ with respect to any vector field $\vec{V}$ is exact
\be
\begin{aligned}\label{eq:cartanid}
\lie_{\vec{V}}\Omega_0&=i_{\vec{V}}d\Omega_0+d\big(i_{\vec{V}}\Omega_0\big)\\
&=d\big(i_{\vec{V}}\Omega_0\big)\\
&=d\tilde{V}.
\end{aligned}
\ee
Here $i_{\vec{V}}\tilde{\omega}$ is the interior product, which contracts $\vec{V}$ with the first entry of any differential form it acts on. In the last line of equation (\ref{eq:cartanid}) we used the zeroth order symplectic form to lower the index $V_B=V^A\Omega^0_{AB}$. Using identity (\ref{eq:cartanid}),  equation (\ref{eq:1stcond}) becomes
\be
d\tilde{\zeta}_1
=-d\tilde{\xi}^{[1]} .
\ee
From the definition of $\tilde{\xi}$ in (\ref{eq:defxi}) we obtain the solution
\be\label{eq:zeta1sol}
\begin{aligned}
\zeta_1^A=&\Omega_0^{AB}\xi^{[1]}_B\\
=&-\frac{1}{2}\Omega^{AB}_0\sum_{n=2}^N  \Big[\frac{\partial}{\partial Q_1^B}K^{(1)}_n(Q_1,\dots,Q_n)\Big]_{\{Q_j\}=Q}.
\end{aligned}
\ee
Now, we use the identity (\ref{eq:cartanid}) in equation (\ref{eq:2ndcond}) to get
\be
\begin{aligned}
d \tilde{\zeta}_2&=-d\tilde{\xi}^{[2]}+\frac{1}{2}\mathscr{L}_{\vec{\zeta}_1}d\tilde{\zeta}_1\\
&=-d\tilde{\xi}^{[2]}+d\big(\frac{1}{2}\mathscr{L}_{\vec{\zeta}_1}\tilde{\zeta}_1\big).
\end{aligned}
\ee
A solution of this equation for the second order vector field is
\be\label{eq:zeta2sol}
\zeta_{2}^A=\Omega_0^{AB}\xi_{B}^{(\N,2)}-\frac{1}{2}\Omega_0^{AB}\big(\lie_{\vec{\zeta}_1}\tilde{\zeta}_1\big)_B.
\ee
It is easy to see that using equation (\ref{eq:allconds}) and the identity (\ref{eq:cartanid}) and the fact that exterior derivatives and Lie derivatives commute, we can find solutions for the vector fields $\vec{\zeta}_i$ that parametrize the diffeomorphism $\varphi(\epsilon)$ up to any order.

\subsection{Transformed second order Hamiltonian}

We now compute the transformed Hamiltonian function (\ref{eq:transformedhamsys}) starting with the expression (\ref{eq:Hdef}) for the $\N^{\text{th}}$ order Hamiltonian $H^{(\N)}$ and specializing to second order for simplicity. The second order Hamiltonian will be expressed in terms of following functions
\begin{subequations}
\begin{align}\label{eq:2nddefphi}
\Phi^{(2)}(Q)&=\Phi(Q,Q,[\bar{X}^{(1)}]),\\ \label{eq:2nddefpsi}
\Psi^{(2)}(Q)&=\Psi(Q,[\bar{X}^{(1)}])
\end{align}
\end{subequations}
where the right hand side terms were defined in equations (\ref{eq:phi}), (\ref{eq:defpsi0}) and (\ref{eq:defpsi}). Both $\Phi^{(2)}$ and $\Psi^{(2)}$ in equations (\ref{eq:2nddefphi}) and (\ref{eq:2nddefpsi}) have contributions of order $O(\epsilon)$ and $O(\epsilon^2)$.

We now specialize the order of the expansion of the diffeomorphism (\ref{eq:diffeo}) to second order. Its action on the Hamiltonian will produce a new Hamiltonian $\hat{H}^{(2)}=\varphi_* H^{(2)}$ given by
\be
\hat{H}^{(2)}=\left(1+\epsilon\lie_{\vec{\zeta}_1}+\epsilon^2\lie_{\vec{\zeta}_2}+\frac{1}{2}\epsilon^2\lie_{\vec{\zeta}_1}\lie_{\vec{\zeta}_1}\right)H^{(2)}+O(\epsilon^3).
\ee
We can simplify this expression using the results (\ref{eq:zeta1sol}) and (\ref{eq:zeta2sol}) for $\vec{\zeta}_1$ and $\vec{\zeta}_2$. We can also use equation (\ref{eq:1formorders}) to regroup $\epsilon \xi^A_{[1]}+\epsilon^2 \xi^A_{[2]}=\xi^A_{(2)}+O(\epsilon^3)$. The result is
\be\label{eq:newham}
\begin{aligned}
\hat{H}^{(2)}&=H^{(2)}-\xi^A_{(2)}\partial_A H^{(2)}-\frac{1}{2}\Omega_0^{AB}\left(\lie_{\vec{\xi}_{(2)}}\tilde{\xi}_{(2)}\right)_B \partial_A  H^{(2)} \\
&+\frac{1}{2}\xi^A_{(2)}\partial_A \left(\xi^B_{(2)}\partial_B H^{(2)}\right) + O(\epsilon^3 ).   
\end{aligned}
\ee
%
In order to calculate $\hat{H}$ we'll make frequent use of the identity (\ref{eq:step5}), specialized to $\N=2$, which becomes
\be\label{eq:iddeltaom}
\Delta \Omega^{(2)}_{AB}\left.\frac{d\bar{X}^{(1)B}_s}{ds}\right|_{s=0}=\left[\frac{\partial}{\partial Q^{A'}} \Phi(Q,Q',[\bar{X}^{(1)}])\right]_{Q'=Q}+\partial_A \Psi^{(2)}(Q)+O(\epsilon_n^{3}).
\ee
We'll also use 
\be\label{eq:idxi1}
\xi_A^{(2)} \frac{d\bar{X}^{(1)A}}{ds}=\frac{1}{2}\Phi^{(2)}+\Psi^{(2)}
\ee
which can be derived from equation (\ref{eq:zeta1sol}) using techniques similar to the ones in subsection \ref{sec:localderiv} (See, for example, equation (\ref{eq:step0})). 

The first correction in equation (\ref{eq:newham}) is $\xi_{(2)}^A \partial_A H^{(2)}$. We use the equations of motion (\ref{eq:hameqs1}) to replace $\partial_A H^{(2)}$ by $\Omega^{(2)}_{AB}d\bar{X}^{(1)B}/ds$
\be
\xi_{(2)}^A \partial_A H^{(2)}=\xi_{(2)}^A \big(\Omega^0_{AB}+\Delta\Omega_{AB}^{(2)}\big)\frac{d\bar{X}^{(1)B}}{ds}+O(\epsilon^3)
\ee
Now, we use identity (\ref{eq:idxi1}) for the first term and identity (\ref{eq:iddeltaom}) for the second term to get
\be\label{eq:newhamstep1}
\begin{aligned}
\xi^A_{(2)}H^{(2)}&=\frac{1}{2}\Phi^{(2)}+\Psi^{(2)}\\
&+\xi_{(2)}^A\left[\frac{\partial}{\partial Q^{A'}}\Phi(Q,Q',[\bar{X}^{(1)}])\right]_{Q'=Q}+\xi_{(2)}^A\partial_A \Psi^{(2)}+O(\epsilon^3).
\end{aligned}
\ee
The second correction term in equation (\ref{eq:newham}) is more involved, let's simplify it first. Using Cartan's magic formula we can write
\be\label{eq:xilieofxi}
\begin{aligned}
\lie_{\vec{\xi}_{(2)}}\tilde{\xi}_{(2)}&=i_{\vec{\xi}_{(2)}}d\tilde{\xi}_{(2)} \\
&=i_{\vec{\xi}_{(2)}}\Delta \Omega^{(2)}
\end{aligned}
\ee
where we used equation (\ref{eq:symisexact}) to replace $d\tilde{\xi}_{(2)}$ by the correction to the symplectic form $\Delta \Omega^{(2)}$. Next, we use the equations of motion to replace $\partial_A H^{(2)}$ by $\Omega^0_{AB}\frac{d\bar{X}^{(1)B}}{ds}+O(\epsilon^2)$. Combining this with equation (\ref{eq:xilieofxi}), the second term in equation (\ref{eq:newham}) becomes
\be
-\frac{1}{2}\Omega^{AB}_0 \xi_{(2)}^C \Delta \Omega_{CB}\Omega_{AD}^0 \frac{d\bar{X}^{(1)D}}{ds}.
\ee
Now, we use identity (\ref{eq:iddeltaom}) to get
\be\label{eq:newhamstep2}
\frac{1}{2}\xi^A_{(2)}\left[\frac{\partial}{\partial Q^{A'}} \Phi(Q,Q',[\bar{X}^{(1)}])\right]_{Q'=Q}+\frac{1}{2}\xi^A_{(2)}\partial_A \Psi^{(2)}(Q)+O(\epsilon_n^{3}).
\ee
The last term in equation (\ref{eq:newham}) is 
\be
\frac{1}{2}\xi_{(2)}^A\partial_A\big(\xi_{(2)}^B \partial_B H^{(2)}\big).
\ee
Again, we use Hamilton's equations to replace $\partial_B H^{(2)}=\Omega^0_{BC}\frac{\bar{X}^{(1)C}}{ds}+O(\epsilon^2)$. We then use identity (\ref{eq:idxi1}) to get
\be\label{eq:newhamstep3}
\frac{1}{2}\xi_{(2)}^A\partial_A \big(\frac{1}{2}\Phi^{(2)}+\Psi^{(2)}\big).
\ee
Combining equations (\ref{eq:newhamstep1}), (\ref{eq:newhamstep2}) and (\ref{eq:newhamstep3}) and plugging them into equation (\ref{eq:newham}), the final expression for the new Hamiltonian is
\begin{equation}
\begin{aligned}
    \hat{H}^{(2)}&=H_0+\frac{1}{2}\Phi^{(2)}-\frac{1}{4}\xi_{(2)}^A\partial_A \Phi^{(2)}\\
    &+\frac{1}{2}\xi_{(2)}^A\left[\frac{\partial}{\partial Q^{A}}\Phi(Q,Q',[\bar{X}^{(1)}])\right]_{Q'=Q}+ O(\epsilon^3).
\end{aligned}
\end{equation}
Note that the third and fourth terms include contributions of order $O(\epsilon^3)$ which could be discarded without affecting the accuracy of the result. 

\section{Application: Hamiltonian description of the conservative self-force on point particles in general relativity}\label{sec:sf}

We now turn to studying binary systems in the small mass-ratio approximation in general relativity. These systems consist of a primary object of mass $M$ and a secondary of mass $m$ with $m\ll M$ orbiting around it. The dynamics of the secondary are described by its position and momentum $Q^A=(x^\mu,p_\mu)$. When the secondary's mass is zero, it moves on a geodesic determined by the metric sourced by the primary. To leading order in the mass ratio $\epsilon\equiv m/M$, the motion of the secondary deviates from geodesic motion due to its interaction with its own gravitational field, known as the self-force. In \cite{Blanco:2022mgd}, we found that the conservative piece of the gravitational self-force to leading order in the mass ratio can be derived from a non-local action principle with zeroth order Hamiltonian
\be
H_0(Q)=\sqrt{-g^{\mu\nu}p_\mu p_\nu}
\ee
and a non-local perturbation
\be
S_{\text{nl}}[X]=\frac{\epsilon}{2} \int dsds' G\big[\bar{X}^{(1)}_s(Q),\bar{X}^{(1)}_{s'}(Q)\big]+O(\epsilon^2)
\ee
where the 2-point function is
\begin{equation}
G(Q,Q')=G^{\mu\nu\alpha'\beta'}(x,x')\frac{p_\mu p_\nu p_{\alpha'}p_{\beta'}}{\big(-g^{\rho\sigma}p_\rho p_\sigma\big)\big(-g^{\rho'\sigma'}p_{\rho'}p_{\sigma'}\big)}.
\end{equation}
Here $G^{\mu\nu\alpha'\beta'}(x,x')$ is the time symmetric Green's function for the linearized Einstein equations in the Lorenz gauge. The parameter $s$ is proper time in the background metric. The leading order conservative piece of the scalar and electromagnetic self-forces can be derived from the same Hamiltonian by replacing the 2-point function by
\begin{subequations}
    \begin{align}
G_{\text{scalar}}&(Q,Q')=G(x,x'),\\
G_{\text{EM}}&(Q,Q')=G^{\mu\nu'}(x,x')\frac{p_\mu p_{\nu'}}{\sqrt{-g^{\rho\sigma}p_\rho p_\sigma}\sqrt{-g^{\rho'\sigma'}p_{\rho'} p_{\sigma'}}}
    \end{align}
\end{subequations}
with $G$ and $G^{\mu\nu}$ the time-symmetric pieces of the Green's function for the Klein-Gordon equation and the Maxwell equations, respectively. The gravitational, electromagnetic and scalar Green's functions are regularized using the Detweiler-Whiting prescription \cite{Detweiler:2002mi}. Since these 2-point functions are all symmetric under exchange of arguments, the results of section \ref{sec:localham} show that the conservative first order dynamics have a local Hamiltonian description. This was shown in \cite{Blanco:2022mgd}, using a method more restrictive than the one presented in this paper, valid only to first order in perturbation theory. 

In \cite{Blanco:2024}, we show how to express the second-order self-force as the integral of a 3-point function and then apply the results of this paper to derive the Hamiltonian description of the conservative piece of the scalar self-force up to second order for nonspinning particles in any stationary spacetime.

\section{Application: Binary systems in general relativity in the post-Newtonian approximation}\label{sec:pn}

The motion of binary systems in general relativity can be studied in the post-Newtonian approximation, where their dynamics is expanded in powers of $1/c^2$. A term of order $1/c^{2n}$ is called $n$PN in the literature. In \cite{4PNDamour2014}, Damour, Jaranowski and Schäfer give an explicit expression for the 4PN non-local Hamiltonian\footnote{In a follow-up paper \cite{localpseudohamDamour2015}, the same authors utilize an (infinite-)order-reduction of the nonlocal
dynamics to a local dynamical system. This procedure is similar to the one carried in section \ref{sec:orderred} and, similarly, doesn't result in a Hamiltonian system.  Instead, the procedure determines a pseudo-Hamiltonian dynamical system (see appendix \ref{sec:appendix2} for details). } of two non-spinning point particles with phase space coordinates $Q^A=(\mathbf{x}_a,\mathbf{p}_a)$ and masses $m_a$ with $a=1,2$ and boldface representing 3-vectors. Following the notation of this paper, we use $X_s$ for a trajectory in phase space parametrized by $s$. Their result is 
\be\label{eq:ham4PN}
H_{\leq \text{4PN}}(Q,[X])=H_{<\text{4PN}}(Q)+H_{\text{4PN}}^{\text{local}}(Q)+H_{\text{4PN}}^{\text{non-local}}(Q,[X])
\ee
where $H_{<\text{4PN}}$ gathers all the contributions of order 3PN or less and $H_{\text{4PN}}^{\text{local}}$ gives the local piece of the 4PN Hamiltonian. We'll focus on the last term, which is written in terms of the quadrupole moment
\be
I^{ij}(\mathbf{x}_a)=\sum_{a=1}^2 m_a\Big( x^i_a x^j_a -\frac{1}{3}\delta^{ij}|\mathbf{x}_a|^2\Big)
\ee
as
\be\label{eq:hamPN}
H_{\text{4PN}}^{\text{non-local}}(Q,[X])=\frac{1}{c^8}\mathcal{C} \dddot{I}^{ij}(Q)\int_{-\infty}^{\infty} d \tau \frac{\dddot{I}_{ij}\ \big(X_\tau\big)}{|\tau|}
\ee
where $\mathcal{C}$ is a normalization factor whose value is not important here. $H_{\text{4PN}}^{\text{non-local}}$ is the non-local or "tail" piece of the 4PN Hamiltonian. The non-locality arises from the integral over the full trajectory $X_\tau$. In equation (5.1) of \cite{4PNDamour2014}, they also derive a non-local contribution to the action principle from which $H_{\text{4PN}}^{\text{non-local}}$ can be derived, which is
\be\label{eq:actionPN}
S_{\text{nl}}[X]=-\frac{1}{c^8}\mathcal{C}\int d\tau d\tau' \frac{\dddot{I}^{ij}\big(X_\tau\big)\dddot{I}_{ij}\big(X_{\tau'}\big)}{|\tau-\tau'|}.
\ee
Note that in equations (\ref{eq:hamPN}) and (\ref{eq:actionPN}) we are dropping the regularization prescription used in \cite{Sch_fer_2018} to take care of the ultraviolet divergences of $H_{\text{4PN}}^{\text{non-local}}$ that occur at the coincidence limit $\tau\rightarrow\tau'$. The regularization can be reapplied after a local Hamiltonian is obtained. 

We now show that the dynamical system (\ref{eq:ham4PN}) can be casted as a local Hamiltonian system by using the results of \ref{sec:localham}. We define a two-point function 
\be
\mathcal{G}_2(Q_1,Q_2,\sigma)=\mathcal{C}\frac{\dddot{I}^{ij}\big(Q_1\big)\dddot{I}_{ij}\big(Q_2\big)}{|\sigma|}
\ee
such that the non-local action in (\ref{eq:actionPN}) takes the form of equation (\ref{eq:nonlocalaction}).  Following the steps of section \ref{sec:localham}, we can evaluate the functional dependence of the non-local Hamiltonian (\ref{eq:hamPN}) on the 0PN flow $\bar{X}^{(0)}$, which is the Newtonian solution to the equations of motion. It is not necessary to include corrections of order $1/c^2$ or higher in the flow, since that would give corrections to the Hamiltonian at 5PN and higher. 

It follows that the non-local Hamiltonian $H_{\leq \text{4PN}}$ admits a local Hamiltonian description up to $O(1/c^8)$, with Hamiltonian function and symplectic form given by the results in section (\ref{sec:localham}).

\section{Conclusions}
In this paper we described a class of dynamical systems whose equations of motion are derived from non-local action principles. We reviewed the well known procedure for deriving local equations of motion by treating the non-localities perturbatively. Then we proved that the perturbative local dynamics admit a local Hamiltonian descriptions up to any order in perturbation theory. We discussed a diffeomorphism on phase space that puts the symplectic form into canonical form up to any order and gave an explicit expression for the new Hamiltonian up to second order in perturbation theory. Finally, we applied these results to the small mass-ratio and post-Newtonian approximations for the study of binary systems in the context of general relativity.
\newline

\vspace{0.1cm}
\noindent \textit{Acknowledgments}: We thank Eanna Flanagan for helpful discussion and comments. We also thank Mohammed Khalil for clarifying the application of the results of this paper to the post-Newtonian approximation.

\color{black}


\twocolumngrid
\bibliography{Ref.bib}

\appendix
\section{Relation to the work of Llosa and Vives}\label{sec:appendix}

Llosa and Vives \cite{llosa1994hamiltonian} consider non-local-in-time action principles in configuration space $(x^\mu,\dot{x}^\mu)$, which they describe as a non-local Lagrangian $L[x]$ which is a functional of $x$. In this paper we consider, instead, an action functional of a phase space trajectory $Q^A=(x^\mu,p_\mu)$. Furthermore, they don't carry the perturbative expansion of the non-localities explicitly but rather leave the non-local piece of the action principle unspecified. This affects their final results in two ways. First, their expressions for the local Hamiltonian and symplectic forms depend on functional derivatives of the action functional. Second, without using a perturbative expansion of the non-localities, the space of initial data for the Hamiltonian flow is not defined. They assume that an order reduction procedure to make the dynamics local exists and work with this unspecified space of initial data instead.  In this paper, we expand the non-local-in-time piece of the action functional as a series of integrals of N-point functions, which allows us to derive a simpler local Hamiltonian and symplectic form, expressed explicitly in terms of integrals of said N-point functions, evaluated on points in the unperturbed phase space, which constitutes our space of initial data. Although it is possible that the results of section \ref{sec:localham} could be obtained from results in their work, our results are derived using a different method and provide a simpler and more streamlined framework for studying non-local-in-time perturbations to all orders. Sections \ref{sec:diffeo}, \ref{sec:sf} and \ref{sec:pn} are entirely original results.
\section{Relation to pseudo-Hamiltonian systems}\label{sec:appendix2}

We define a {\it pseudo-Hamiltonian} dynamical system to consist of a phase space $\Gamma$, a closed,
non-degenerate two form $\Omega_{AB}$ and a smooth pseudo-Hamiltonian function
${\cal H} : \Gamma \times \Gamma \to {\mathbb R}$, for which the dynamics
are given by integral curves of the vector field
\be
v^A =  \Omega^{AB} \frac{\partial}{\partial Q^B} \left. {\cal H}(Q,Q')
\right|_{Q'=Q},
\label{phdef}
\ee
where $\Omega^{AB} \Omega_{BC} = \delta^A_C$ and $Q^A$ are coordinates on $\Gamma$.

The perturbative local dynamical systems derived in subsection \ref{sec:orderred} are examples of pseudo-Hamiltonian systems which are perturbations of a Hamiltonian system. The symplectic form and pseudo-Hamiltonian up to $\N^{\text{th}}$ order are
\begin{subequations}
  \label{phpfull}
\begin{eqnarray}
  \Omega_{AB} &=& \Omega_{0\,AB},\\
        {\cal H}^{(\N)}(Q,Q') &=& H_0(Q) + \Phi(Q,Q',[\bar{X}^{(\N-1)}])\label{php},
\end{eqnarray}
\end{subequations}
where $\Phi(Q,Q',[X])$ is defined in equation (\ref{eq:phi}). The local equations of motion (\ref{eq:hameqs1}) are obtained by plugging the pseudo-Hamiltonian system (\ref{phpfull}) into equation (\ref{phdef}).

In this paper, we derived pseudo-Hamiltonian equations of motion from a non-local action principle. However, pseudo-Hamiltonians can be used in a broader context, and need not be derived from a variational principle. In that case, the n-point functions $\mathcal{G}_n$ that appear in the definition (\ref{eq:phi}) of $\Phi(Q,Q',[X])$ need not satisfy the symmetry property (\ref{eq:symprop}). A pseudo-Hamiltonian system obtained by starting from equations (\ref{php}) and (\ref{eq:phi}), without imposing that the n-point functions obey the symmetry property (\ref{eq:symprop}) can include dissipative effects  \cite{Galley:2012hx}. In the context of the first order gravitational self force, for example, we can construct a pseudo-Hamiltonian using the retarded Green function, which encodes both dissipative and conservative effects, as opposed to the time-symmetric Green function, which only describes the conservative piece of the dynamics.

\end{document}